\documentclass[12pt]{article}
\usepackage{latexsym}
\usepackage{amsmath,amssymb}
\usepackage{bm}
\usepackage{graphicx,color}
\usepackage{amssymb}
\usepackage{cite}
\usepackage{amsfonts, amscd, amsthm}
\usepackage[all]{xy}
\usepackage{here}
\usepackage{ulem}
\usepackage{mathabx}
\usepackage[titletoc,title]{appendix}
\usepackage{mathbbol}

\renewcommand{\theequation}{\thesection.\arabic{equation}}

\makeatletter
\@addtoreset{equation}{section}
\renewcommand{\theequation}{\thesection.\@arabic\c@equation}
\makeatother

\makeatletter
\renewcommand\appendix{\par
  \setcounter{section}{0}%
  \setcounter{subsection}{0}%
  \gdef\thesection{Appendix \@Alph\c@section }
  \renewcommand{\theequation}
  {\Alph{section}.\arabic{equation}}
}
\makeatother

\makeatletter
\newcounter{subeqncnt}
\def\thesubeqncnt{\alph{subeqncnt}}
\def\subequations{\begingroup%
\stepcounter{equation}\edef\@tempa{\theequation}%
\let\c@equation\c@subeqncnt\c@subeqncnt\z@
\edef\theequation{\@tempa\noexpand\thesubeqncnt}}

\makeatother

\allowdisplaybreaks
\begin{document}
\titlepage

\title{Various Logistic Curves in SIS and SIR Models}

\author{
Kazuyasu Shigemoto\\
Tezukayama University, Nara 631-8501, Japan\\
}
\date{\empty}

\date{\empty}

\maketitle

\abstract{ In our previous paper, the logistic curve of the removed number was derived from SIR and SEIR 
models in the case of the small basic reproduction number. 
In this paper, we derive various logistic curves of the removed,  unsusceptible and infectious numbers
respectively from SIS and SIR models in the case of small and large  basic reproduction numbers.
}\\

\hskip 10mm
\noindent
[ {\bf Keywords}: Various Logistic Curves, SIS model, SIR model, Epidemiology ]

\section{Introduction}
\setcounter{equation}{0}
The logistic growth curve is introduced by Verhulst~\cite{Verhulst} to explain the 
realistic population growth curve instead of the Malthus' exponential growth curve.
Next comes the predator-prey model or Lotka-Volterra equations~\cite{Lotka,Volterra} 
in the general biology. For the special case of the predator-prey model, the logistic curve
emerges.
Then, if we apply this special predator-prey model to the infectious disease, we obtain 
the SIS model by identifying susceptible number and infectious number in the infectious disease as 
prey and predator respectively. Thus, the logistic curve come out in the infectious disease model.
The SIS model is useful to analyze some infectious disease such as 
norovirus disease. In the norovirus disease, even if we infected and recovered, 
we again become susceptible. 

For the infectious disease, where the immunity is acquired  after the recovery,    
 Kermack and McKendrick introduce the 
SIR model~\cite{Kermack} by adding recovered number to the SIS model.
This SIR model is used to analyze the general infectious disease
~\cite{Murray,Hethcote,Brauer,Keeling,Pathak} and the 
recent COVID 19~\cite{Sameni, Saito1,Saito2,Saito3,Saito4,Saito5,Saito6,Saito7}.
Next comes the SEIR model~\cite{Li} to analyze more details of the infectious disease.

In the context of the logistic curve,
the SIS model is the origin of the logistic curve in SIR model.
In the SIS model, the equation becomes the exact 
logistic equation, and the solution gives the exact logistic 
curve. In the SIR model, there emerge various approximate 
logistic equations, which has the origin in the SIS model.

Though equations of the SIR model are the simple coupled differential equations, it is the black box
to solve the model numerically. The SIR is one of the models for the infectious 
disease and about 10$\%$ error is all right for the practical usage. For example, if we compare
results of the SIR model with those of the SEIR model with the same reproduction number, there are 
more than 10$\%$ differences. Then,  the approximate but simple analytic expression 
is quite useful for the practical analysis. We can understand the meaning of the 
result of the infectious disease by using such approximate but simple expression.  
For the logistic curve, we easily know some property 
such as  the place of the inflection point and that the time derivative of that curve becomes 
the left-right symmetric pulse type curve etc..

In the previous paper, we derived the approximate logistic curve of the removed number  for the 
SIR model~\cite{Saito1} and that for the SEIR model~\cite{Saito7}.
In this paper, we derive approximate various logistic curves for 
the removed,  unsusceptible and infectious numbers respectively
in SIS and SIR models in the region of small and large reproduction numbers.
In section 2,  we review SIS model, where we exactly obtain the logistic curve.
In section 3, we derive various approximate logistic curves for the SIR model in the region 
of small and large reproduction numbers.
The final section is devoted to the summary and discussions. 

\section{The SIS model }
\setcounter{equation}{0}
By applying the  predator-prey model or 
Lotka-Volterra equations~\cite{Lotka,Volterra} to the infectious disease model, the SIS model is 
introduced.
By using the prey $u(t)$ and the predator $v(t)$, 
the predator-prey model is given by the coupled differential equation in the form
\begin{eqnarray}
&& \frac{d u(t)}{d t}=a u(t) -b u(t) v(t) ,
\label{2e1}\\
&& \frac{d v(t)}{d t}=c u(t) v(t) -d v(t).  
\label{2e2}
\end{eqnarray}
In the case of $a=d$, $b=c$, we have the conservation law
$\displaystyle{\frac{d(u(t)+v(t))}{dt}=0}$; thus predator-prey model 
becomes the SIS model.

By using the susceptible number $S(t)$ and the infectious number $I(t)$,
the SIS model is given by
\begin{eqnarray}
&& \frac{d S(t)}{d t}=-\beta S(t) I(t) +\gamma I(t) ,
\label{2e3}\\
&& \frac{d I(t)}{d t}=\beta S(t) I(t) -\gamma I(t) .  
\label{2e4}
\end{eqnarray}
where $\beta$ is the infection rate, $\gamma$ is the removed rate and $\alpha=\beta/\gamma$
is the basic reproduction number. Using the conservation law,
$\displaystyle{\frac{d (S(t)+I(t))}{dt}=0}$, we normalized them in the form
\begin{equation}
S(t)+I(t)=1 .
\label{2e5}
\end{equation}
Substituting $I(t)=1-S(t)$ into Eq.(\ref{2e1}), we obtain the logistic  
equation of $S(t)$ in the form
\begin{eqnarray}
&&\frac{d I(t)}{d t}=A_1 I(t)\left(1-\frac{I(t)}{B_1} \right), \quad A_1=\gamma(\alpha-1), \quad
B_1=\frac{\alpha-1}{\alpha}. 
\label{2e6}
\end{eqnarray}
The solution is given by the logistic curve in the form
\begin{eqnarray}
 I(t)=\frac{B_1}{1+\exp( -A_1(t-T)) }, \quad T=\text{const.} .
\label{2e7}
\end{eqnarray}
One of the properties of the logistic differential equation is the followings:
If we take the inverse of the original logistic function in the form $f(t)=1/I(t)$, we obtain 
the linear differential equation in the form
$$\frac{df(t)}{dt}=-A_1 f(t)+\frac{A_1}{B_1}.$$

We define the unsusceptible number $N(t)$ at $t$ by
\begin{eqnarray}
 N(t)=1-S(t). 
\label{2e8}
\end{eqnarray}
In this SIS model, it happens to be $N(t)=I(t)$ and the equation of $N(t)=I(t)$ becomes the 
same form as that of Eq.(\ref{2e4}) in the form
\begin{eqnarray}
&&\frac{d N(t)}{d t}=A_1 I(t)\left(1-\frac{N(t)}{B_1} \right), \quad A_1=\gamma(\alpha-1), \quad
B_1=\frac{\alpha-1}{\alpha}. 
\label{2e9}
\end{eqnarray}
The solution is given by the logistic curve in the form
\begin{eqnarray}
N(t)=\frac{B_1}{1+\exp( -A_1(t-T)) }, \quad T=\text{const.} .
\label{2e10}
\end{eqnarray}
Two logistic equations Eq.(\ref{2e6}) and Eq.(\ref{2e9}),
which become exact equations and they takes the same form,
are the origin that there are various approximate logistic 
equations in the next SIR model.

\section{The SIR model }
\setcounter{equation}{0}
By using  susceptible number $S(t)$, the infectious number $I(t)$ and the removed number $R(t)$,
the SIR equations are given by
\begin{eqnarray}
&& \frac{d S(t)}{d t}=-\beta S(t) I(t) ,
\label{3e1}\\
&& \frac{d I(t)}{d t}=\beta S(t) I(t) -\gamma I(t)  ,
\label{3e2}\\
&& \frac{d R(t)}{d t}=\gamma I(t)   .
\label{3e3}
\end{eqnarray}
where $\beta$ is the infection rate, $\gamma$ is the removed rate and $\alpha=\beta/\gamma$ is the 
basic reproduction number. By using the conservation law 
$\displaystyle{\frac{d(S(t)+I(t)+R(t)}{dt}=0}$, we normalize them in the form
\begin{equation}
S(t)+I(t)+R(t)=1  .
\label{3e4}
\end{equation}
From Eq.(\ref{3e1}) and Eq.(\ref{3e3}), we get $dS(t)/dR(t)=-\alpha S(t)$. 
By using the boundary condition of $R(0)=0$ and $S(0)=1$, we obtain
the solution in the form
\begin{equation}
S(t)=\exp(-\alpha R(t)) .
\label{3e5}
\end{equation}
Then, by using Eq.(\ref{3e4}), we obtain
\begin{eqnarray}
I(t)=1-R(t)-\exp(-\alpha R(t))   .
\label{3e6}
\end{eqnarray}
By substituting this relation into Eq.(\ref{3e3}) we obtain the  equation
of $R(t)$ in the form
\begin{eqnarray}
&&\frac{d R(t)}{dt}=\gamma (1-R(t)-\exp(-\alpha R(t)) )  .
 \label{3e7}
\end{eqnarray}
In the $\alpha=1$ case, Eq.(\ref{3e2}) is written in the form
$d I(t)/d t=\gamma(S(t)-1) I(t) $ with $S(0)=1$, which gives 
$dI(t)/dt|_{t=0}=0$ and $I(t)$ cannot increase from zero, and the solution is given 
by $S(t)=1$, $I(t)=0$, $R(t)=0$.
Hence, we consider only the $(\alpha-1)>0$ region.
In the actual numerical calculation of Eq.(\ref{3e7}), we put the initial condition to be
$R(0)=\epsilon$(=small value) instead of $R(0)=0$. If we put $R(0)=0$, the right-hand side of
Eq.(\ref{3e7}) at $t=0$ becomes $\gamma  (1-R(0)-\exp(-\alpha R(0)) )=0$, which 
gives $dR(t)/dt|_{t=0}=0$ and $R(t)$ does not increase from $0$.

Eqs.(\ref{3e1})-(\ref{3e3}) are numerically solved for various $\alpha$ and $\gamma$.
We give some examples in Figure 1 with $\alpha=1.3$, $\gamma=1$, 
and in Figure 2 with $\alpha=20$, $\gamma=1$.

\begin{figure}[H]
\qquad\quad
 \begin{minipage}{0.4\hsize}
  \begin{center}
\hspace{-10.1mm}
   \includegraphics[width=80mm]{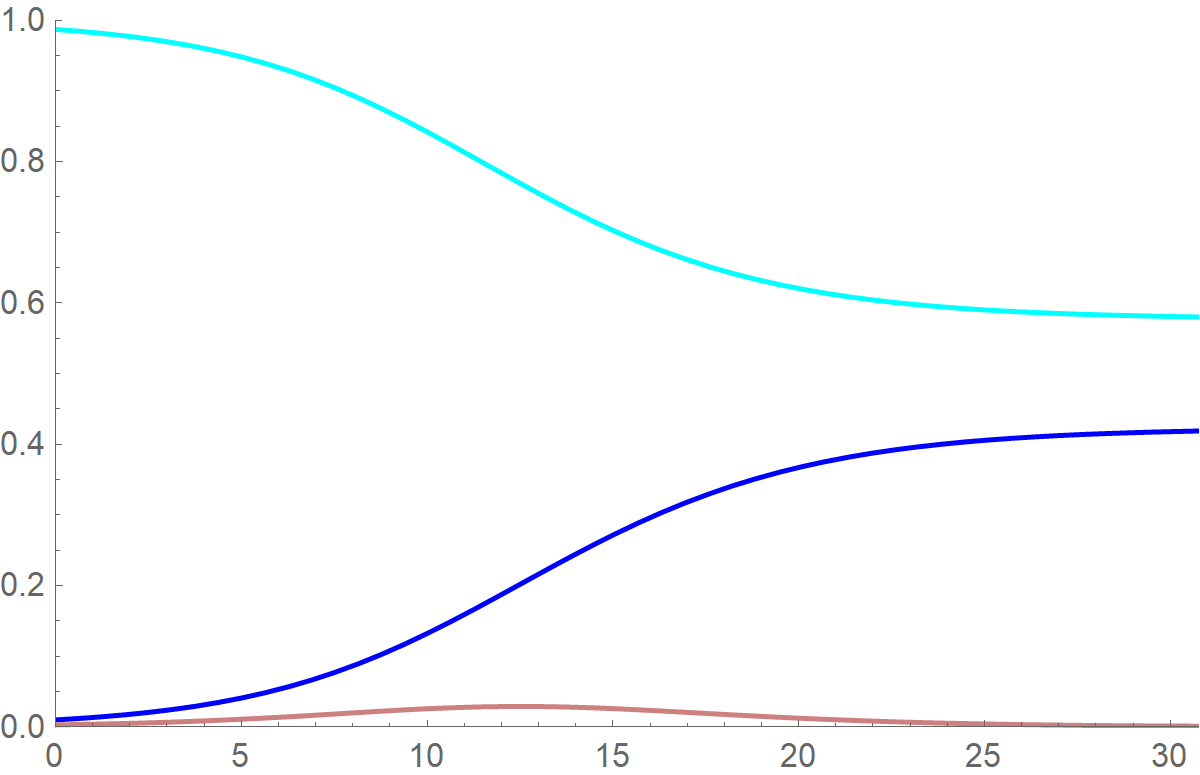}
  \end{center}
 \caption{
$\alpha=1.3$, $\gamma=1$ \, 
\protect\newline
\hspace*{17mm}  
}
  \label{fig1}
 \end{minipage}
 \qquad
 \lower1.3ex\hbox{%
 \begin{minipage}{0.4\hsize}
  \begin{center}
   \includegraphics[width=80mm]{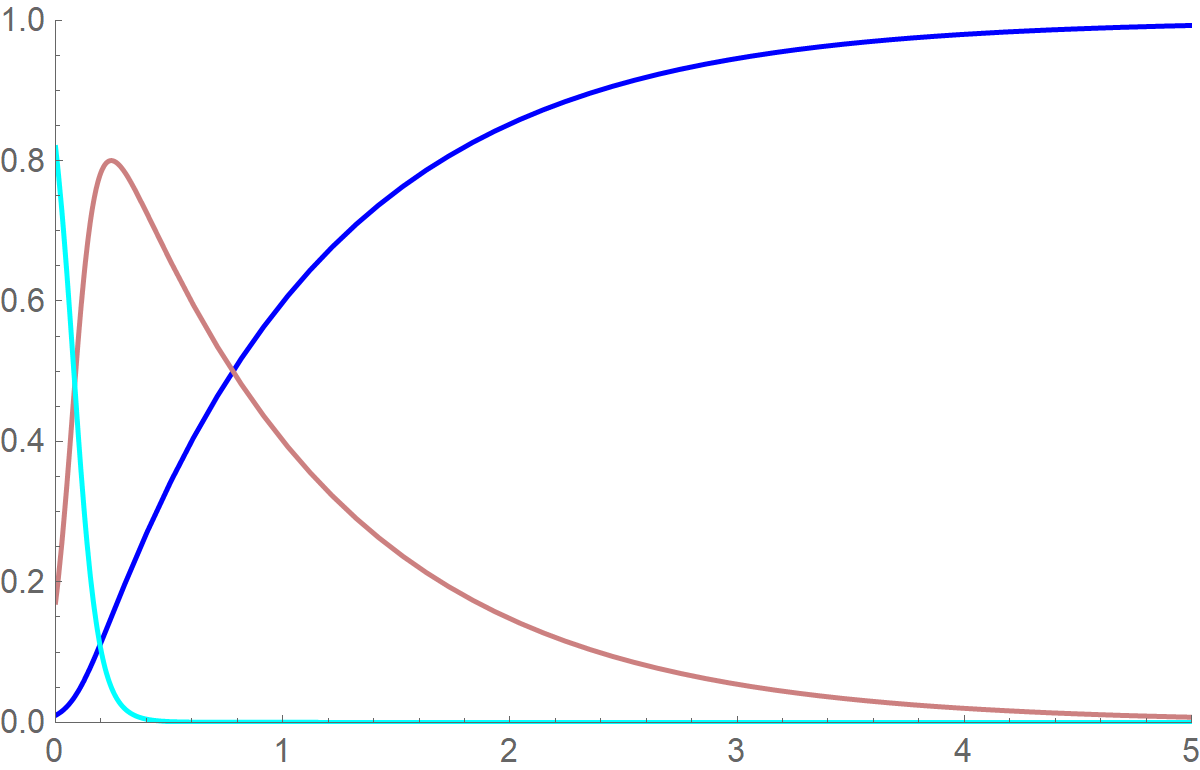}
  \end{center}
  \caption{$\alpha=20$, $\gamma=1$ }
  \label{fig2}
 \end{minipage}
 }
\end{figure}
If $(\alpha-1)$ is small, $S(t)$ remains around 1,  $I(t)$ is quite small and $R(t)$ is rather small. 
As $I(t)$ is quite small, the SIR model
becomes similar to the SIS model, where $R(t)$ in the SIR model corresponds to $I(t)$ in the SIS model. 
If $(\alpha-1)$ is large, $I(t)$ takes the quite high peak, which goes near 1, and $R(t)$
increase up to almost 1 at the end. Before $I(t)$ takes the peak value, $R(t)$ is small. 
Then, before the peak of $I(t)$, SIR becomes similar to the SIS model.
By knowing these properties of the numerical calculation for various $(\alpha-1)$ regions, 
we derive various approximate equations for $R(t)$,$N(t)$ and $I(t)$ in the small $(\alpha-1)$ and
in the large $(\alpha-1)$ cases respectively to find the simple and useful curves.

\subsection{The Logistic Curve of $R(t)$}

\noindent
\underline{{\bf i) $0<(\alpha-1) \ll 1$ case:}}\\
We review the approximate logistic curve of $R(t)$ in the SIR model~\cite{Saito1}.
By using Eq.(\ref{3e6}),  we obtain 
\begin{eqnarray}
I(t)=1-R(t)-\exp(-\alpha R(t)) ,
\label{3e8}
\end{eqnarray}
In this case, we know that $R(t)$ is small from the numerical calculation. 
Then we make the following approximation
\begin{equation} 
\exp(- \alpha R(t)) \approx 1-\alpha R(t)+\frac{\alpha^2 R(t)^2}{2} .
\label{3e9}
\end{equation}
Thus, we obtain the approximate expression of $I(t)$ in the  form
\begin{eqnarray}
I(t) \approx 1-R(t)-(1-\alpha R(t) +\frac{(\alpha R(t))^2}{2})
=(\alpha-1) R(t) \left( 1-\frac{R(t)}{2(\alpha-1)/\alpha^2} \right)   .
\label{3e10}
\end{eqnarray}
Hence, from Eq.(\ref{3e3}), we obtain the logistic equation in the form
\begin{eqnarray}
&&\frac{d R(t)}{dt}=\gamma I(t)=A_2 R(t) \left(1-\frac{R(t)}{B_2}\right), 
\quad A_2=\gamma(\alpha-1), B_2=\frac{2(\alpha-1)}{\alpha^2} .
 \label{3e11}
\end{eqnarray}
The solution is given by the logistic curve in the form
\begin{eqnarray}
 R(t)=\frac{B_2}{1+\exp( -A_2(t-T)) }, \quad T=\text{const.}  .
\label{3e12}
\end{eqnarray}
%

\noindent
\underline{{\bf ii) $(\alpha-1) \gg 1$ case:}}\\
In the $(\alpha-1) \gg 1$ region,  we know that $R(t)$ increases up to almost 1.
As $(\alpha-1)$ is large, $\exp(-\alpha R(t))$ is small compared with $(1-R(t))$ except in the
region of quite small $R(t)$ in such  a way as 
$\alpha R(t) =\mathcal{O}(1)$ in Eq.(\ref{3e8}). In such case, Eq.(\ref{3e7}) gives
\begin{eqnarray}
&&\frac{d R(t)}{dt}-\gamma \left(1-R(t)\right)=0    ,
 \label{3e13}
\end{eqnarray}
which gives the solution
\begin{eqnarray}
 R(t)=1-\exp( -\gamma (t-\epsilon)), \quad \epsilon=\text{small const.} .
\label{3e14}
\end{eqnarray}
Here we put the initial condition of $R(t)$ to be 
$R(0)=1-\exp(\gamma \epsilon)\approx -\gamma \epsilon$ instead of $R(0)=0$.
We can choose such initial condition because we cannot apply Eq.(\ref{3e13}) in the region of 
$R(t) \approx \mathcal{O}(\epsilon)$ as we excluded the region $\alpha R(t) =\mathcal{O}(1)$ from the beginning.

In the large $t$ region, we have 
$ \displaystyle{R(t)=1-\exp( -\gamma (t-\epsilon)) \approx \frac{1}{1+\exp( -\gamma (t-\epsilon))}}$.
Thus, strictly speaking, this $R(t)$ is not the logistic curve, but we also call 
this $R(t)$ as the "logistic" curve here.

\subsection{The Logistic Curve of $N(t)$}
Next, we derive the approximate logistic curve of $N(t)$ in the SIR model.
From the relation $S(t)=\exp{(-\alpha R(t))}$, we obtain $R(t)=-\log S(t)/\alpha$.
By substituting this relation into Eq.(\ref{3e8}), we obtain
\begin{equation}
I(t)=1-S(t)+(\log S(t))/\alpha    .
\label{3e15}
\end{equation}
Then Eq.(\ref{3e1}) becomes in the form
\begin{eqnarray}
\frac{d S(t)}{dt}=-\gamma \alpha S(t) \left(1-S(t)+\frac{\log S(t)}{\alpha} \right)   .
\label{3e16}
\end{eqnarray}
The unsusceptible number $N(t)$ at $t$ is defined by 
\begin{equation}
N(t)=1-S(t)   ,
\label{3e17}
\end{equation}
Thus, Eq.(\ref{3e16}) is given in the form
\begin{eqnarray}
\frac{d N(t)}{dt}
=\gamma \alpha (1-N(t) )\left(N(t)+\frac{\log (1-N(t))}{\alpha} \right)    .
\label{3e18}
\end{eqnarray}

\noindent
\underline{{i) $0<(\alpha-1) \ll 1$ case:}}\\
In this case, we know that $N(t)$ is small ($S(t)$ is almost 1) from the numerical calculation. 
Then we make the following approximation 
\begin{equation}
\log (1-N(t)) \approx -N(t)-\frac{N(t)^2}{2}  .
\label{3e19}
\end{equation}
Thus, Eq.(\ref{3e16}) gives
\begin{eqnarray}
&&\frac{d N(t)}{dt} \approx \gamma \alpha (1-N(t) )
\left((1-\frac{1}{\alpha}) N(t)-\frac{N(t)^2}{2\alpha} \right)=
\gamma(\alpha-1) N(t)(1-N(t) )\left(1-\frac{1}{2(\alpha-1)} N(t) \right) 
\nonumber\\
&& \approx \gamma (\alpha-1)N(t) \left(1-\frac{2\alpha-1}{2\alpha-2} N(t) \right)  .
\label{3e20}
\end{eqnarray}
Hence, we obtain
\begin{eqnarray}
\frac{d N(t)}{dt}=A_3 N(t) \left(1-\frac{N(t)}{B_3}  \right),
\quad A_3=\gamma(\alpha-1), B_3=\frac{2\alpha-2}{2\alpha-1}  , 
\label{3e21}
\end{eqnarray}
The solution is given by the logistic curve in the form
\begin{eqnarray}
N(t)=\frac{B_3}{1+\exp(-A_3(t-T))}, \quad T=\text{const.}  .
\label{3e22}
\end{eqnarray}
%

\noindent
\underline{{\bf ii) $(\alpha-1) \gg 1$ case:}}\\
In this case, we know from the numerical calculation that
$I(t)$ takes the quite high peak and goes near 1, and 
$R(t)$ is small before the peak of $R(t)$.
By using $I(t)=N(t)-R(t)$, the above implies that
$N(t) \approx 1$ and $R(t) \ll 1$ near the peak.
Thus, we have $N(t) \gg R(t)=|\log (1-N(t))/\alpha|$ in this case.
Then Eq.(\ref{3e18}) gives the logistic equation
\begin{eqnarray}
\frac{d N(t)}{dt}
=\beta N(t) (1-N(t) )=A_4 N(t) \left(1-N(t)\right), \quad A_4=\gamma \alpha .
\label{3e23}
\end{eqnarray}
The solution is given by the logistic curve in the form
\begin{eqnarray}
N(t)=\frac{1}{1+\exp(-A_4(t-T))},  \quad T=\text{const.}  .
\label{3e24}
\end{eqnarray}
%

\subsection{The Logistic Curve of $I(t)$}
Further, we derive the approximate logistic curve of $I(t)$ in the SIR model.
By using $I(t)=1-R(t)-S(t)=1-R(t)-\exp(-\alpha R(t))$,  we obtain 
\begin{eqnarray}
&&\frac{d I(t)}{dt}=(-1+\alpha \exp( -\alpha R(t)) )\frac{d R(t)}{dt}
=(-1+\alpha \exp(-\alpha R(t)) )\gamma I(t)
\nonumber\\
&&=\gamma I(t)\Big(-1+\alpha (1-R(t)-I(t)) \Big)
=\gamma I(t)\Big( (\alpha-1)-\alpha(I(t)+R(t))\Big)  .
\label{3e25}
\end{eqnarray}
%

\noindent
\underline{{\bf i) $0<(\alpha-1) \ll 1$ case:}}\\
In this case, we know that $R(t)$ is small. Using $I(t)=1-R(t)-\exp(-\alpha R(t))$, we make the following approximation
\begin{equation}
I(t) \approx 1-R(t)-(1-\alpha R(t)) 
=(\alpha-1) R(t) 
\label{3e26}
\end{equation}
Then we obtain $R(t)=I(t)/(\alpha-1)$ ($I(t)$ is quite small and $R(t)$ is small).
Substituting this relation into Eq.(\ref{3e25}), we obtain 
\begin{eqnarray}
&&\frac{d I(t)}{dt}\approx \gamma I(t)\Big( (\alpha-1)-\alpha I(t) -\frac{\alpha}{\alpha-1}I(t))\Big)
=\gamma I(t)\Big( (\alpha-1)-\frac{\alpha^2}{\alpha-1} I(t) \Big) \nonumber\\
&&=\gamma(\alpha-1) I(t)\Big(1-\frac{\alpha^2}{(\alpha-1)^2} I(t) \Big)   .
\label{3e27}
\end{eqnarray}
Thus, we obtain the logistic equation of the form
\begin{eqnarray}
&&\frac{d I(t)}{dt}=A_5 I(t) \left(1-\frac{I(t)}{B_5} \right), \quad A_5=\gamma(\alpha-1),
B_5=\frac{(\alpha-1)^2}{\alpha^2}    . 
\label{3e28}
\end{eqnarray}
The solution is given by the logistic curve in the form
\begin{eqnarray}
I(t)=\frac{B_5}{1+\exp(-A_5(t-T))}, \quad T=\text{const.}  .
\label{3e29}
\end{eqnarray}

\noindent
\underline{{\bf ii) $(\alpha-1) \gg 1$ case and $R(t) \ll I(t)$ region :}}\\
In this case and region, we know from the numerical calculation that 
$I(t)$ takes the high peak and goes near 1. In the region before the peak of $I(t)$, $R(t)$ remains small.
Then Eq.(\ref{3e25}) gives 
\begin{eqnarray}
&&\frac{d I(t)}{dt}
\approx \gamma (\alpha-1) I(t) \Big( 1-\frac{\alpha}{\alpha-1}I(t) \Big)
\label{3e29}
\end{eqnarray}
Thus, we obtain the logistic equation
\begin{eqnarray}
&&\frac{d I(t)}{dt}
=A_6 I(t) \left(1-\frac{I(t)}{B_6} \right), \quad A_6=\gamma(\alpha-1),
B_6=\frac{\alpha-1}{\alpha}  .
\label{3e30}
\end{eqnarray}
The solution is given by the logistic curve in the form 
\begin{eqnarray}
I(t)=\frac{B_6}{1+\exp(-A_6(t-T))}, \quad T=\text{const.}  .
\label{3e31}
\end{eqnarray}
%

Below, we give the graph in the small $(\alpha-1)$ approximation. Figure 3 is the $\alpha=2.5$ case and 
Figure 4 is the $\alpha=10$ case in the small $(\alpha-1)$ approximation.
\begin{figure}[H]
\qquad\quad
 \begin{minipage}{0.4\hsize}
  \begin{center}
\hspace{-10.1mm}
   \includegraphics[width=80mm]{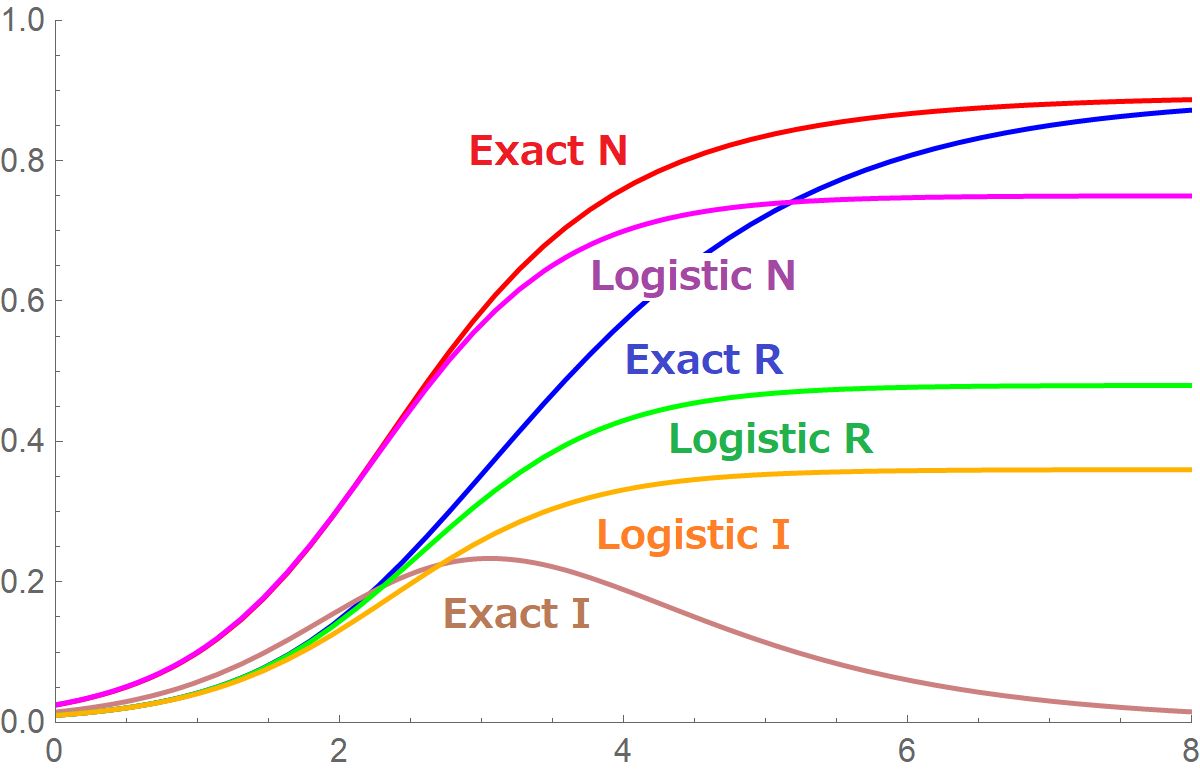}
  \end{center}
 \caption{Small $(\alpha-1)$ approximation \
\protect\newline
\hspace*{17mm}  $\alpha=2.5$, $\gamma=1$
}
  \label{fig3}
 \end{minipage}
 \qquad
 \lower1.3ex\hbox{%
 \begin{minipage}{0.4\hsize}
  \begin{center}
   \includegraphics[width=80mm]{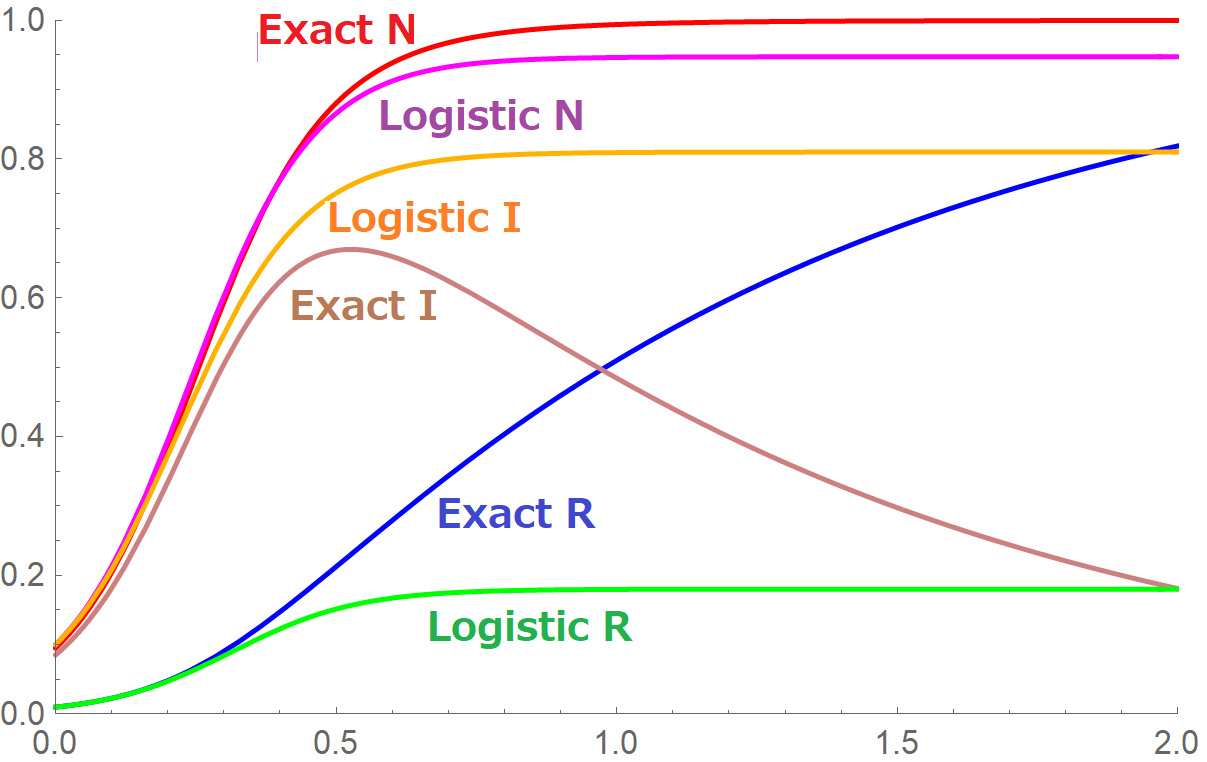}
  \end{center}
  \caption{Small $(\alpha-1)$ approximation \
\protect\newline
\hspace*{17mm}  $\alpha=10$, $\gamma=1$
}
  \label{fig4}
 \end{minipage}
 }
\end{figure}
Below, we give the graph in the large $(\alpha-1)$ approximation. Figure 5 is the $\alpha=20$ case and 
Figure 6 is the $\alpha=5$ case in the large $(\alpha-1)$ approximation.
\begin{figure}[H]
\qquad\quad
 \begin{minipage}{0.4\hsize}
  \begin{center}
\hspace{-10.1mm}
   \includegraphics[width=80mm]{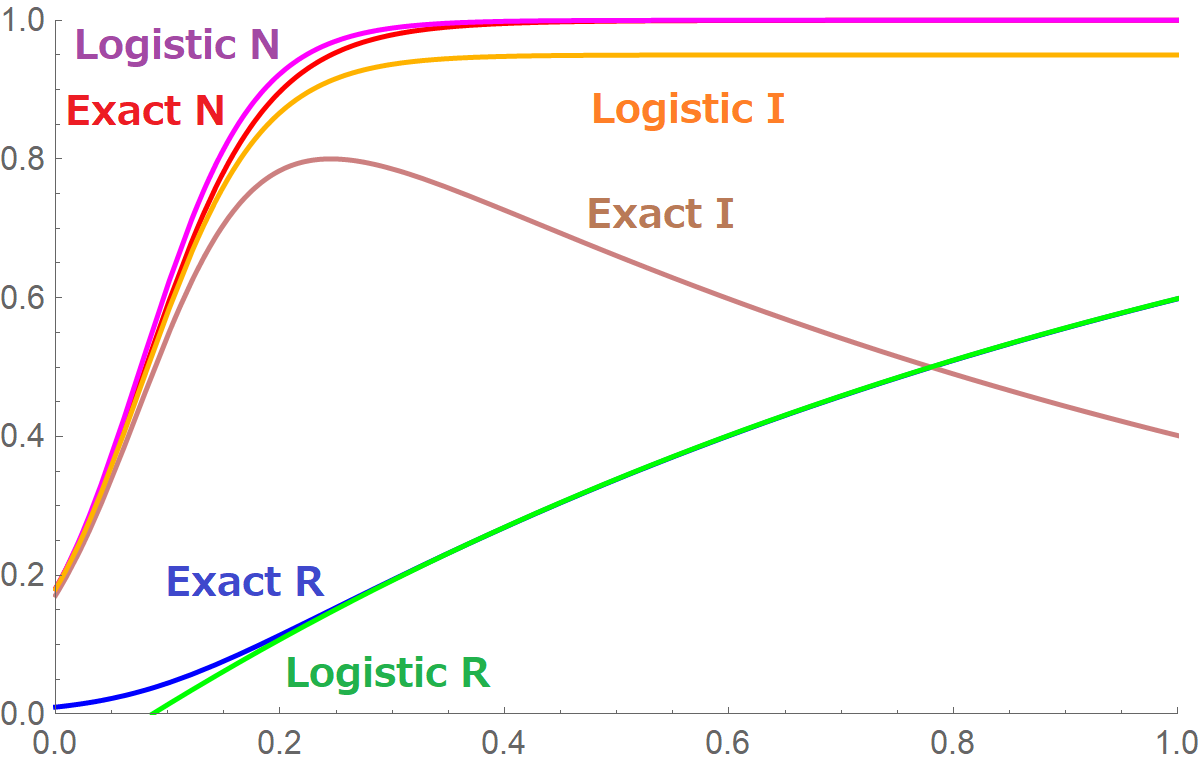}
  \end{center}
 \caption{Large $(\alpha-1)$ approximation \ 
\protect\newline
\hspace*{17mm}  $\alpha=20$, $\gamma=1$
}
  \label{fig5}
 \end{minipage}
 \qquad
 \lower1.3ex\hbox{%
 \begin{minipage}{0.4\hsize}
  \begin{center}
   \includegraphics[width=80mm]{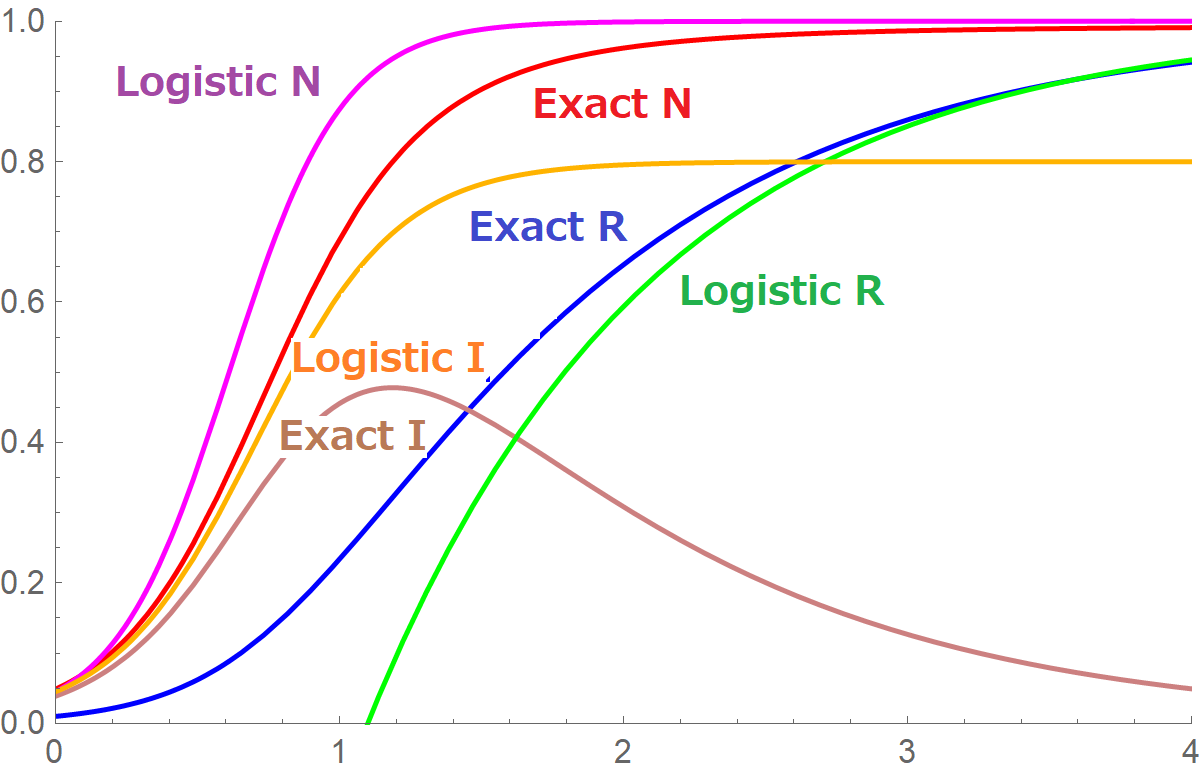}
  \end{center}
  \caption{ Large $(\alpha-1)$ approximation \
\protect\newline
\hspace*{17mm}  $\alpha=5$, $\gamma=1$
}
  \label{fig6}
 \end{minipage}
 }
\end{figure}

\section{Summary and Discussions}
\setcounter{equation}{0}
We first review the logistic curve in SIS model. In the SIS model, the  equations of $N(t)$ and $I(t)$ 
becomes the exact logistic  equation, and we obtain the exact logistic solution.
In the SIR model, we derived various approximate  logistic equations of $R(t)$, $N(t)$ 
and $I(t)$ in small $(\alpha-1)$
and large $(\alpha-1)$ regions. We obtained the common feature 
that $A_1=A_2=A_3=A_5=A_6=\gamma(\alpha-1)$
and $A_4=\gamma \alpha$ in the $(\alpha-1)\gg 1$ case, that is, $A_n (n=1,2,\cdots,6)$ take essentially  
the common value.  Only the difference is the final value $B_n (n=1,2, \cdots, 6)$ at $t \rightarrow \infty$.
Thus, we obtain any logistic curve from one logistic curve by compressing or expanding in the vertical 
direction except the "logistic" curve $R(t)$ in the large $(\alpha-1)$ approximation.

For the small $(\alpha-1)$ approximation, in $\alpha=2.5$ case, the obtained approximate 
logistic curve of $N(t)$ is applicable 
to the most long time interval. Next is the logistic curve of $R(t)$ and the logistic curve of 
$I(t)$ can be used up to the peak of exact $I(t)$. 
In $\alpha=10$ case, though it is the result of small $(\alpha-1)$ approximation, the obtained approximate 
logistic curve of $N(t)$ fits well for the long time interval.
The logistic curve of $I(t)$ fit well before the peak of exact $I(t)$. The logistic curve of $R(t)$ does not fit well 
at the region of large time because logistic $R(t)$ in this approximation has the property 
$B_2 \rightarrow 0$ as $\alpha \rightarrow \infty$.

For the large $(\alpha-1)$ approximation, in $\alpha=20$ case, the obtained approximate logistic 
curve of $N(t)$ and 
$R(t)$  fits well for the long time interval. The logistic curve of $I(t)$ fits rather well 
before the peak of exact $I(t)$. 
In $\alpha=5$ case,  the obtained approximate logistic curve of $N(t)$ fits well, 
$I(t)$ curve fits rather well before the peak of exact $I(t)$, 
and $R(t)$ curve fits only at the region of large time.

\section*{Acknowledgement}
I would like to express my deep gratitude to Prof. T. Saito for many valuable discussions.


\end{document}